\documentclass[aps,prd,twocolumn,showpacs,floatfix,superscriptaddress,
preprintnumbers,nofootinbib]{revtex4}
\usepackage{graphicx,amssymb}

\newcommand{\be}{\begin{equation}}
\newcommand{\ee}{\end{equation}}
\newcommand{\ba}{\begin{eqnarray}}
\newcommand{\ea}{\end{eqnarray}}

\def\lsim{\raise0.3ex\hbox{$\;<$\kern-0.75em\raise-1.1ex\hbox{$\sim\;$}}}
\def\gsim{\raise0.3ex\hbox{$\;>$\kern-0.75em\raise-1.1ex\hbox{$\sim\;$}}}
\def\eps{\varepsilon}

\begin{document}

\title{Annihilations of superheavy dark matter in superdense clumps}

\author{V.~Berezinsky}
 \affiliation{INFN, Laboratori Nazionali del Gran Sasso, I--67010
  Assergi (AQ), Italy}
 \affiliation{Center for Astroparticle Physics at LNGS (CFA), I--67010
  Assergi (AQ), Italy}
 \affiliation{Institute for Nuclear Research of the Russian Academy of
 Sciences, Moscow, Russia}
\author{V.~Dokuchaev}
 \author{Yu.~Eroshenko}
 \affiliation{Center for Astroparticle Physics at LNGS (CFA), I--67010
  Assergi (AQ), Italy}
 \affiliation{Institute for Nuclear Research of the Russian Academy of
 Sciences, Moscow, Russia}
\author{M.~Kachelrie\ss}
 \affiliation{Institutt for fysikk, NTNU Trondheim, N--7491 Trondheim,
  Norway}
\author{M.~Aa.~Solberg}
 \affiliation{Institutt for fysikk, NTNU Trondheim, N--7491 Trondheim,
  Norway}

\date{February 18, 2010}

\begin{abstract}
Superheavy dark matter (SHDM) exchanges energy with its environment much 
slower than particles with masses close to the electroweak (EW) scale  
and has therefore different 
small-scale clustering properties. Using the neutralino as candidate for 
the SHDM, we find that free-streaming allows the formation of DM
clumps of all masses down to $\sim 260\,m_\chi$ in the case of bino. 
If small-scale clumps evolve from a non-standard, spiky spectrum of 
perturbations, DM clumps may form during the radiation dominated era. 
These clumps are not destroyed by tidal interactions and can be extremely 
dense. In the case of a bino, a ``gravithermal catastrophe'' can develop in 
the central part of the most dense clumps, increasing further the central density
and thus the annihilation signal. In the case of a higgsino, the annihilation 
signal is enhanced by the Sommerfeld effect. As a result 
annihilations of superheavy neutralinos in dense clumps may lead to observable 
fluxes of annihilation products in the form of ultrahigh energy particles, 
for both cases,  higgsinos and binos, as lightest supersymmetric particles.  
\end{abstract}

\pacs{
12.60.Jv,
95.35.+d, 
14.80.Nb, 
98.70.Vc, 
95.85.Pw  
}

\maketitle

\section{Introduction}

The case for the existence of non-relativistic, non-baryonic dark matter
(DM) in the universe is stronger than ever~\cite{DMreviews}. But although
a wealth of observational data provides compelling evidence for a $\sim23\%$
contribution of DM  to the total energy density of the universe, its nature 
is still not known. The most popular DM type are thermal relics, i.e.\ 
particles that were at least once during the history of the Universe in 
chemical equilibrium with the thermal plasma.

The present relic abundance $\Omega_\chi$ of a thermal relic scales 
approximately with its annihilation cross section $\sigma_{\rm ann}$ as 
$\Omega_\chi\propto 1/\sigma_{\rm ann}$. Moreover, unitarity bounds 
annihilations as $\sigma_{\rm ann}\propto m_\chi^{-2}$ and thus the
observed value~\cite{WMAP5} $\Omega_{\rm CDM}h^2=0.1$ of the DM
abundance constrains the annihilation cross section as
$\langle\sigma_{\rm ann}v\rangle\sim 3\times 10^{-26}$cm$^3$/s and
limits the mass of any thermal relic as $m_\chi\lsim 50\,$TeV~\cite{GKH}. 
Hence thermal relics offer detection prospects both for direct and 
indirect searches as well as at accelerator experiments.

The assumption that the DM particle was once in chemical
equilibrium is however not necessary and does not hold in
particular for sufficiently heavy particles. Superheavy particles
are generated at the end of inflation and they can play the role
of DM particles~\cite{BKV,KR}. Gravitational production at the end 
of inflation provides the most natural mechanism for the generation
of superheavy dark matter (SHDM)~\cite{grav}. 
Their decays can result in the observable signal in the form of 
UHE gamma-rays~\cite{BKV} and UHE neutrinos~\cite{BKmc},
if they are metastable and long-lived.

While the idea of SHDM is theoretically appealing, the detection of 
SHDM is challenging. Clearly,
accelerator searches and direct detection are not feasible in the
case of SHDM. The feasibility of indirect detection of {\em stable} 
SHDM depends on its annihilation rate $\dot N_{\rm ann}\propto
(\rho/m_\chi)^2 \langle\sigma_{\rm ann}v\rangle$ that in turn
scales naively as $\dot N_{\rm ann}\propto m_\chi^{-4}$. Since backgrounds
like cosmic rays from astrophysical sources or the diffuse photon
flux decrease only as  $1/E^{\alpha}$ with $\alpha\lsim 3$,
indirect detection of DM seems to become more and more difficult
for increasing DM masses. The only possibility which overcomes this 
difficulty is the superdense central region of DM clumps \cite{BlasiDickKolb02}, but
one needs the realistic scenario for the very high density of DM in the clump center or
formation of superdense clumps \cite{kt,SS}.  

We shall use as candidate for SHDM the neutralino in the model
of superheavy supersymmetry, as suggested in Ref.~\cite{BKS}.
Superheavy supersymmetry is a unique scheme that respects perturbative
unitarity despite of coupling particles with mass much larger than
the weak scale to the electroweak sector. Within this model, the
lightest supersymmetric particle (LSP) that we choose as the neutralino
is a natural candidate for SHDM.

Aim of the present work is to study the detection prospects for
{\em stable\/} SHDM. Since the annihilation signal from the mean
distribution of DM in the halo is far below observational limits,
we examine if new effects specific for SHDM exist that can improve
the detection chances. 

One such effect can result from the early
kinetic decoupling of SHDM: While the mass spectrum of DM clumps formed
by standard neutralinos with mass in the 100\,GeV range has a
cutoff at $M_{\rm min}\sim (10^{-12}-10^{-4})M_{\odot}$~\cite{bde06,cutweak}, 
the cutoff can be  diminished significantly e.g.\ in the case of 
ultra-cold WIMPs~\cite{GelGon08,SS}. 
We will show that the cutoff is practically absent for
a superheavy bino as DM particle, and clumps of all masses are possible
beginning formally from $\sim 260\,m_\chi$. This low-mass cutoff
increases the diffuse flux of UHE particles produced by annihilations.

Another effect is the formation of superdense clumps, in which the 
annihilation rate can be strongly enhanced. The formation of
superdense clumps is discussed in general in the accompanying paper 
I~\cite{pI}. 
Here we study this problem in more detail for superheavy particles. 

The initial mass spectrum of DM clumps is determined by the
spectrum of cosmological density perturbations. The simplest models
of inflation  predict a nearly scale-invariant  power-law form
that is then normalized to the COBE observations. Clumps formed in
this case are not very dense, and the small-scale structure of DM
enhances the annihilation signal not strongly. A quite different
scenario arises, if the potential
$V(\phi,T)$ of the inflaton field contains features like a zero
derivative $V'(\phi,T)$ for some field value $\phi$. The spectrum
of perturbations in this case has a spike and  clumps of SHDM can
form already in the radiation-dominated (RD) epoch, and 
they can have very large densities. 

For sufficiently dense clumps, the relaxation processes due to 
gravitational two-body scatterings 
can initiate a ``gravithermal catastrophe'' and the density profile of 
the clump
core steepens to an isothermal profile $\rho\propto r^{-2}$ with a much 
smaller new core radius. While this process was discussed in general 
in Ref.~\cite{pI} (hereafter Paper I), we show in this work
that for a bino this process can take place and determine the required
initial conditions. Annihilations of DM in such dense clumps are strongly 
amplified because of the density enhancement.

In contrast to a bino, winos and higgsinos are stronger coupled to the
thermal plasma. As a result, a ``gravithermal catastrophe'' does not
develop. However, the velocities of DM particles in superdense clumps
are very small, leading to an enhancement by the Sommerfeld effect in 
case of higgsinos or winos. Taking into account both effects, we find
that annihilations of SHDM bound in such superdense clumps
may lead to observable fluxes of ultrahigh energy particles for all types
of neutralinos.

Note that (in contrast to the case with standard power-law
spectrum of cosmological perturbations) superdense clumps produced
in the RD epoch from a spiky spectrum are not destroyed by tidal forces and their
mass function peaks near a definite value. Therefore the
fraction of DM in the form of such clumps is $\xi\simeq1/2$. 

This article is organized as follows. We start with a discussion of the
energy relaxation time 
of SHLSPs in Sec.~\ref{sec:cross}, where we also derive  the minimal
mass of the SHDM clumps
considering free-streaming and the horizon scale at kinetic decoupling.
In Sec.~\ref{sec:clump} we summarize the evolution of the density profile
of superdense clumps. 
Next we discuss the prospects to detect SHDM clumps in Sec.~\ref{sec:detect}
and conclude in Sec.~\ref{sec:conc}.

We shall use below the following abbreviations: SHDM for superheavy 
dark matter and SHLSP for superheavy lightest supersymmetric 
which in our case is the superheavy neutralino~\cite{BKS}. We fix
the mass $m_\chi$ of the SHLSP as $m_\chi=10^{11}$\,GeV, i.e.\ within the
range suggested by gravitational production with $\Omega_\chi h^2\approx 0.1$
at the end of inflation.

\section{Energy relaxation and the minimal mass of DM clumps}
\label{sec:cross}
\newcommand{\mc}{\mathcal}

\subsection{Relaxation time}

We shall consider here the energy relaxation of SHDM particles
interacting with a thermal background at temperature $T$. The
consideration is relevant for both ordinary and superdense clumps. 

The energy exchange between superheavy neutralinos and light fermions 
was calculated in \cite{BKS} for temperatures $T$ below the weak scale, i.e.\ 
in the limit $T\ll m_Z  \ll M_{\rm SUSY}$ with $M_{\rm SUSY}=\min\{M_1,M_2,\mu\}$.
Here, $\mu$ denotes the Higgs mixing parameter, and $M_1$ and $M_2$ the U(1)
and SU(2) soft SUSY breaking masses. At lowest order in $m_Z^2/M_{\rm SUSY}^2$,
the neutralino masses  are simply $\{ M_1, M_2, -\mu, \mu \}$, for more
details see e.g.\ \cite{BKS}.

The energy relaxation time is calculated as
\be   \label{E:enreltimedef}
  \tau_{\rm rel}^{-1} =
  \frac{1}{2E_k m_{\chi}}\sum_i\int_0^\infty \!\!\!\!\!d\omega \int d\Omega\:
  n_i(\omega)(\delta p)^2 \left(\frac{d\sigma_{{\rm el},i}}{d\Omega}\right) \, ,
\ee
where $E_k=(3/2)T$ is the mean kinetic energy of the neutralinos,
$\delta p$ the neutralino momentum obtained in one scattering,
\be
  (\delta p)^2=2\omega^2[1-\cos(\theta)]
\ee
and the number density of relativistic fermions or bosons with $g$
polarization degrees and energy $\omega$ is
 \be
  n_i=\frac{g_i}{2\pi^2}\frac{\omega^2}{e^{\omega/T}\pm 1}\approx
  \frac{g_i}{2\pi^2}\:\omega^2 \, e^{-\omega/T} \,.
 \ee
Kinetic decoupling occurs when the  energy relaxation rate
$\tau_{\rm rel}^{-1}$ becomes smaller than the expansion rate $H$
of the universe. During the radiation dominated (RD) epoch, $H=1/(2t)$,
and
\be
 H = 1.66 \sqrt{g_\ast} \: \frac{T^2}{M_{\rm Pl}} \,,
\ee
where $g_{\ast}$ denotes the number of relativistic degrees of freedom
and $M_{\rm Pl}$ the Planck mass, $M_{\rm Pl}=1/\sqrt{G_N}\simeq
1.2\times 10^{19}\,$GeV.

As the calculations of  Ref.~\cite{BKS} show 
the energy-relaxation time due to elastic scattering of
neutralino on the background fermions at $T \lsim m_Z$  
is larger than the corresponding age of the universe. Thus,
it is  necessary to  include also the elastic scattering of 
neutralino on
weak gauge bosons and the light higgs for $m_Z\ll T\ll M_{\rm SUSY}$.
Since the mass splitting between the lightest neutralino and chargino
is of order ${\cal O}(m_Z^2/M_{\rm SUSY})$ for a higgsino or wino, inelastic
processes like $\chi^0+\nu_e\to \chi^++e^-$ contribute also to the energy
exchange between neutralinos and the plasma. As we will
see, elastic interactions with fermions are for $T\gg m_Z$ subdominant,
while inelastic ones give the dominant contribution to the energy
relaxation of the neutralino. Since the latter are absent for a bino,
the relaxation time is strongly dependent on the nature of the neutralino.
We consider in the following only the cases of a bino and a higgsino,
noting that a wino behaves in most respects as a higgsino.

\subsubsection{The Bino as the LSP}

The processes  dominating the energy exchange between superheavy binos and
the thermal
plasma at $T\gg m_Z$ are $\chi Z \to \chi Z$, $\chi W^\pm \to \chi W^\pm$
and $\chi h \to \chi h$. In the first process only light higgs exchange
contributes at leading order and the squared amplitude is
\be
|\mc{M}_{\chi Z \to \chi Z}|^2=
\frac{e^4 M_1^2 \left(\mu \sin(2\beta) + M_1\right)^2}
     {3\cos^4(\theta_W)(\mu^2-M_1^2)^2} \,,
\ee
where $\tan\beta=v_1/v_2$ is the ratio of the two Higgs vev  and
$\theta_W$ the Weinberg angle.
The matrix elements of the other relevant processes have the same form, and
the total matrix element squared weighted by the relevant polarization degrees
is given by $|\mc{M}|^2=\sum_i g_i |\mc{M}_i|^2=12|\mc{M}_{\chi Z \to \chi Z}|^2$.
The resulting energy relaxation time follows as
\be
 \tau_{\rm rel} =
 \frac{\pi c_W^4 M_1 (\mu^2-M_1^2)^2}
      {32\alpha^2 T^4 (M_1+\mu s_{2\beta})^2} \sim
      \frac{A\:M_{\rm SUSY}^3}{T^4} \,.
\ee
In the last step, we simplified the energy relaxation time $\tau_{\rm rel}$
introducing a common mass scale $M_{\rm SUSY}$ and $A=\pi c_W^4/(32\alpha^2)$.

\subsubsection{The Higgsino as the LSP }

The inelastic process $\chi^0_1 \nu_e \to \chi^+_1 e^-$ is dominated by
$W$ exchange. Its leading contribution to the squared  matrix element
is given by
\be
 |\mc{M}_{\chi^0_1\nu_e\to \chi^+_1 e^-}|^2 =
 \frac{2 e^4 \mu^2 \omega^2 \cos^2\left(\theta/2\right)}
      {s_W^4\left(2 \omega ^2 (1-\cos(\theta ))+m_W^2\right)^2 } \,.
\ee

The energy relaxation time becomes for $\mu\gg T\gg m_W$
%
\be
 \tau_{\rm rel} = \frac{6\pi \sin^4\left(\theta_W\right)\mu}
                       {\alpha^2 N_{\text{eff}} T^2
                  \left[\ln \left(4 T^2/m_W^2\right)-2 \gamma +1\right]}
\ee
and
\be
 \tau_{\rm rel} \sim \frac{B\mu m_W^4}{N_{\text{eff}} T^6} 
\qquad{\rm with}\quad
 B = \frac{\pi \sin^4\left(\theta_W\right)}
          {160 \alpha^2}
 \ee
for $m_{\chi^\pm}-m_{\chi^0}\ll T\ll m_W$. The factor $N_{\rm eff}$ counts
the number of fermions contributing to the inelastic processes at temperature
$T$. 

\subsection{Minimum mass of DM clumps}
\label{mmins}

The growth of small-scale fluctuations can be smeared out by destructive
processes such as free-streaming, acoustic oscillations and 
others~\cite{cutweak,LoeZal05} 
(see appendix~A in Ref.~\cite{bde08} for a general discussion). 
For neutralinos with mass close to the electroweak (EW) scale, these 
processes determine the minimal wave-length
in the perturbation spectrum that can grow and thereby also the minimum clump
mass. Here we consider SHLSPs and study the effectiveness of these damping
effects in the case $m_\chi\gg m_Z$.

Let us calculate the cosmological age $t_d$ and the temperature
$T_d$ of kinetic decoupling of SHLSPs from the cosmic plasma, i.e.\ the
moment $t_d$ when the relaxation rate $\tau^{-1}_{\rm rel}$ equals 
the expansion rate $H(t_d)$ of the Universe. In
evaluating the condition for decoupling, $\tau_{\rm rel}^{-1}\simeq H$,
we set $N_{\rm eff}=431/4=107.75$ as number of relativistic degrees in
the standard model (SM) for $T>m_t$ and use for the running of coupling
and mixing parameters with temperature $T$ the SM relations,
$\sin^2\theta_W(T)=1/6+5\alpha(T)/[9\alpha_s(T))]$.
For $M_{\rm SUSY}=10^{12}\,$GeV we obtain then
\be          
 \label{tdtd}
 T_d \simeq
\left\{ 
  \begin{array}{ll}
        2\times10^{11} \,\mbox{GeV} \,, & \quad {\rm bino} \\
        2  \,\mbox{GeV} \,, & \quad {\rm higgsino}
  \end{array}
\right. \,,
\ee
as decoupling temperature for a bino and higgsino, respectively.
In the former case, $g_\ast=298/4$ and $N_{\rm eff}=66$.
For $t>t_d$, the SHLSP is not longer in thermal equilibrium with
the cosmic plasma and its momentum scales as $p\propto 1/a^2$.

We consider now the physical processes relevant for a spherical region 
containing DM with total mass $M$ close to the time of horizon crossing. 
The mass of DM inside the horizon as function of the temperature is given by
\begin{equation}
 M=3.4\times10^{16}(T/100\mbox{GeV})^{-3}(N_{\rm eff}/100)^{-3/4}\mbox{~g}.
\label{mdmt}
\end{equation}
In particular, at the temperature of kinetic decoupling given by
Eq.~(\ref{tdtd}) the corresponding mass is equal to
\be           \label{md2}
 M_d \simeq
\left\{ \begin{array}{ll}
        6\times10^{-12} \,\mbox{g} \,, & \quad {\rm bino} \\
        6\times10^{21} \,\mbox{g} \,, & \quad {\rm higgsino}
        \end{array}
\right. \,.
\ee
For a bino, the mass $M_d$ is only 34 times greater than the particle
mass $m_{\chi}\sim10^{11}$~GeV$=1.78\times10^{-13}$~g.

The evolution of fluctuations with mass $M\ll M_d$ and $M\gg M_d$ is
rather different after horizon crossing~\cite{LoeZal05}.
Fluctuations in DM with mass $M\ll M_d$ run out as sound waves in
the radiation plasma. These fluctuations do not have a kick in the peculiar
velocities of their DM particles and therefore do not grow
logarithmically. After kinetic decoupling their amplitude freezes in
until the matter dominated (MD) epoch, and their evolution is analogous
to the evolution of entropy perturbations and described by the
Meszaros solution (see e.g. \cite{kt}). Therefore, there is a  steepening 
in the mass spectrum below  $M\sim M_d$.

In the opposite case, $M\gg M_d$, the peculiar velocities just after
the horizon crossing are equal to $v_{pH}\simeq\delta_Hc/3$
\cite{ind}. In contrast to thermal velocities these peculiar
velocities are regular and directed toward the center of the
fluctuation. The fluctuations grow according to the adiabatic law
$\delta\propto \ln(t)$ due to the  evolution of the peculiar velocities
as $v_{p}(t)\simeq v_{pH}a(t_H)/a(t)$. The free streaming scale
$\lambda_{\rm fs}$ of SHDM is very small. Expressed in comoving
units, this scale  is growing during the RD epoch as
\begin{equation}       \label{fs}
 \lambda_{\rm fs}(t)=a(t_0)\int_{t_d}^{t}\frac{v(t')dt'}{a(t')}
 = 2t_dv_d\frac{a(t_0)}{a_d}{\rm ln}\frac{a(t)}{a_d} \,,
\end{equation}
where $v(t)=v_da(t_d)/a(t)$, $v_d=(3T_d/m)^{1/2}$, and the ratio of the
scale factors $a(t)/a(t_d)$ is calculated from the Friedmann equation.
The corresponding free streaming mass,
\begin{equation}       \label{Mfs}
M_{\rm fs}(t)=\frac{4\pi}{3}\rho_c(t_0)\Omega_{m,0}\lambda_{\rm fs}^3(t) \,,
\end{equation}
stops growing near the epoch of matter-radiation equality, $t\sim
t_{\rm eq}$. For the case of a bino in Eq.~(\ref{tdtd}), the decoupling time 
is $t_d=7\times10^{-30}$~s, and we find
\ba
 M_{\rm fs} & = &\frac{\pi^{1/4}}{2^{19/4}3^{1/4}}
 \frac{\rho_{\text{\rm eq}}^{1/4}t_d^{3/2}}{G^{3/4}}
 \left(\frac{T_d}{m_{\chi}}\right)^{3/2}\!\!\!\ln^3\left\{\frac{24}{\pi
 G\rho_{\text{\rm eq}}t_d^2}
\right\}
\nonumber\\
& \simeq & 4.6\times10^{-11}\mbox{~g}. \label{mfs} \ea
This value is only 260 times larger the particle mass. Thus the
free-streaming mass of SHDM defines the cutoff in the mass spectrum.
Formally, all clump masses are possible  from 
$M \gsim 260 m_\chi$. 

In the case of a higgsino, $M_{\rm fs}\ll m_\chi$, and  free-streaming 
plays no role for the evolutions of density perturbations.

Therefore, the two mass scales $M_d$ and $M_{\rm fs}$ may play the role of 
$M_{\rm min}$. In the case of a bino, $M_{\rm fs} > M_d$ and the steepening of the 
mass spectrum starts at $M_{\rm min} \sim M_{\rm fs}$. In the case of a higgsino, 
$M_{\rm fs}$ is very small and $M_{\rm min} \sim M_d$. 

\section{Clump structure for standard and spiky perturbations}
\label{sec:clump}

Let us consider first the formation and evolution of clumps of SHDM for a
standard power-law spectrum of fluctuations. 
Clumps of SHDM are produced and evolve according to the usual hierarchical
model, as described in Ref.~\cite{bde03} and Paper I, with the essential 
difference that the minimum clump mass is now the one derived in 
Sec.~\ref{mmins}.
With an accuracy sufficient for this schematic consideration, we can use 
$M_{\rm min}$ of order of $260m_{\chi}$ for the
case of a bino. 

The basic features of this scenario for SHDM clumps are the same as 
for ordinary DM clumps: Most clumps are destroyed by tidal 
interactions in hierarchical structures and the surviving clumps could be
further destructed in the Milky Way (see Paper~I). SHDM clumps in this
scenario have a rather small density and SHDM particles with their small 
annihilation cross section are unobservable through their annihilation
products. They can be detected only gravitationally as discussed in
Paper~I.   

Let us come now to the case of a spiky  perturbation spectrum, or
to  any other case with fluctuations growing in the RD epoch. In these cases
superdense clumps can be produced. Since in the RD dominated epoch  
large-scale structures are absent, there is no tidal destruction of small 
clumps. These clumps evolve as isolated objects without 
any merging. The first stage of evolution, the ordinary gravitational
contraction, proceeds in the standard way leading to a 
$\rho(r) \propto r^{-1.8}$ density profile and a large core with 
radius $R_c \sim (0.01 - 0.1)R$, where $R$ is the clump radius.
At this stage the core size is restricted by tidal forces \cite{bde03}, by 
a decreasing mode of
perturbations \cite{ufn1} or by phase density constraints according to 
Liouville's theorem \cite{IshMakEbi}. 
The ``gravithermal instability'' sets in, when the density of the core 
reaches a critical value that is determined mainly by the mass of the
DM particle, and leads to an isothermal density profile  
$\rho (r) \propto r^{-2}$, with a new small core, determined by the
properties of SHLSP. This so-called ``gravithermal catastrophe'' 
occurs under the influence of two-body gravitational scattering in full
analogy of this process to the one in globular clusters. We will discuss 
this process now in some detail. 

\subsection{Gravitational relaxation and evolution of the clumps}
\label{gravrels}

The gravithermal instability in globular clusters sets in due to 
two-body gravitational relaxation. This process can be the dominant one 
for the superdense clumps from SHDM particles.
The other relaxation channel, EW scattering of superheavy neutralinos
loses the competition, because its cross section is proportional to 
$m_{\chi}^{-2}$ and the interaction has a short range. On the other hand,  
the very high clump density provides the relaxation time $t_{\rm rel,gr}$ 
to be shorter than the age of the universe $t_0$. 

The two-body gravitational relaxation time determined as 
$t_{\rm rel,gr}^{-1} = (1/E)(dE/dt)$ in energy space    
can be taken from calculations for globular clusters  (see \cite{Spit} 
and references therein)  as 
\begin{equation}
 t_{\rm rel,gr}\simeq\frac{1}{4\pi}\frac{v^{3}}{G^{2}m_{\chi}^{2}n\ln(0.4N)},
 \label{trelg}
\end{equation}
where $v\sim(GM/R)^{1/2}$ is the virial velocity, and $N$ is 
the total number of particles in the clump, $N=M/m_{\chi}$. 
From Eq.(\ref{trelg}) one observes indeed that the relaxation time is 
inversely proportional the mass squared $m_{\chi}^2$ of SHDM particle and
the density  $n$ of the core. The logarithmic term $\ln(0.4 N)$ takes into
account the long-range character of the gravitational interaction.

The relaxation time shorter than the age $t_0$ of the
universe leads to the ``gravithermal catastrophe'', which  
results in an isothermal density profile $\rho(r)\propto r^{-2}$.
In this regime the main process responsible for the evolution of the clumps 
becomes the evaporation of particles from the core, and the following 
calculations are in full analogy with the
globular clusters case \cite{Spit}. The escaping particles have 
approximately zero total energy, and therefore the energy of the core 
is approximately constant. The process of the core evolution can be
described by the rate of evaporation and the virial theorem, which 
can be written as
\begin{equation}
\dot N_c/N_c=-a/t_{\rm rel,gr},\qquad \dot R_c/R_c=2\dot
N_c/N_c-\dot E_c/E_c \,,
\label{vir}
\end{equation}
where $a\approx7.4\cdot10^{-3}$ for a Maxwellian initial velocity 
distribution. 
Joint integration with logarithmic accuracy [$\ln(0.4N)={\rm const}$ in
Eq.~(\ref{vir})] and the condition $E_c={\rm const}$ gives the time evolution for the core
mass $M_c$ and radius $R_c$:
\begin{equation}
M_c(t)=m_{\chi}N_c(t)=M_{c,i}(1-(t-t_{i})/t_{e})^{2/7},
\ee
\be
R_c(t)=R_{c,i}(1-(t-t_{i})/t_{e})^{4/7},\label{b2}
\end{equation}
where $t_{e}=2/(7\alpha)t_{{\rm rel,gr},i}\simeq 40t_{{\rm
rel,gr},i}$, and the subscript $i$ marks the values at the initial
moment $t_i$ of the clump formation. The time $t_e$ is less than
the age of the Universe for clumps above the dotted line in
Fig.~\ref{figmulti} for $m_{\chi}=10^{11}$\,GeV as mass of the DM particle. 
Thus for clumps above the
dotted lines, relaxation results in the ``gravithermal catastrophe'' producing an isothermal profile $\rho\propto
r^{-2}$ with a tiny new core. It diminishes  until the central density
becomes  sufficiently large and new processes, such as 
annihilations or the pressure of a degenerate Fermi gas, 
enter the game (see below).

\subsection{Formation of the new core}
\label{elrels}

When the large initial core loses its stability and contracts 
under the two-body gravitational forces, the gravitational relaxation
time is increasing. It is caused by the increase of the core density as 
\be
\rho(t) \propto \left [1-(t-t_{i})/t_{e}\right ]^{-10/7} ,
\label{rho(r)}
\ee
which follows from Eqs.~(\ref{b2}). In principle, this
phenomenon can stop the contraction and stabilize the core. 

After the core collapsed, the singular profile $\rho\propto r^{-2}$ extends 
formally down to some small radius $R_c$. For the density profile 
$\rho(r)=\rho_c (r/R_c)^{-2}$, the relative core radius $x_c=R_c/R$ is given 
by $x_c=(\bar\rho/3\rho_c)^{1/2}$, where $\bar\rho$ and $\rho_c$ are the mean
and the maximal density of the clump. Do any physical 
processes exist that prevent the extremely large densities, i.e.\ a very
small radius, of the new core? 

The first candidate for such process is given by the EW elastic 
scattering of SHDM particles i.e.\ by selfinteraction.
The relaxation time for this process can be estimated as  
\begin{equation}
t_{{\rm rel},\chi\chi}^{-1}\simeq\frac{4(2\pi)^{1/2}v
\sigma_{\chi\chi}n_c}{3^{5/2}},
 \label{relax2}
\end{equation}
where $\sigma_{\chi\chi}$ is the cross section of elastic neutralino scattering 
\begin{equation}
 \frac{d\sigma_{\chi\chi}}{d\Omega}\simeq \frac{A\alpha^2}{m_\chi^2}\,,
\label{sxx1}
\end{equation}
where $A$ is a constant of order one that depends on the SUSY parameters, 
$\alpha=1/137$, and $v\simeq(GM/R)^{1/2}$ is the virial velocity. 
This relaxation time has the same dependence on the core-density 
($\propto n_c$) as the gravitational relaxation time (\ref{trelg}), but 
is for SHLSPs many magnitudes smaller. Therefore, self-interactions 
cannot stop the gravitational collapse.  One may see this effect in a 
different way: The core
remains transparent for superheavy neutralinos down to extremely small 
radii. Moreover, \cite{bergur} proved that  elastic
scatterings do not change the central distribution of DM. 
The effect of elastic relaxations in the models of self-interacting DM 
was studied also, e.g., in \cite{koch00}.

Another effect which can stop the gravitational contraction is SHDM
particle annihilation. This effect was studied in Refs.~\cite{bergur}
and {\cite{berez96}. In the former the core radius was found  
from the condition that the characteristic annihilation time in the core    
should be longer than the time of core formation estimated as hydrodynamical
free-fall time $t_h \sim (G\bar\rho)^{-1/2}$.
For an isothermal density profile, the corresponding dimensionless  core 
radius $x_c$ is given by 
\begin{equation}
x_c^2\simeq\frac{\langle\sigma_{\rm ann}
v\rangle\rho^{1/2}}{G^{1/2}m}.
\end{equation}
The mass of superheavy particle $m$ in denominator makes the core
radius rather  small,
\begin{equation}
x_c\simeq7.4\cdot10^{-13}m_{11}^{-3/2}\left(\frac{\rho}{10^5\mbox{ g
cm$^{-3}$}}\right)^{1/4},\label{annihcore}
\end{equation}
where $m_{11}=m/(10^{11}\mbox{~GeV})$.

In a more reliable approach \cite{berez96}, the radius of the core 
was estimated equating the core accretion rate with the 
annihilation rate inside it. The accretion rate was calculated from
the Euler and Poisson equations. The calculated core radius is  
much smaller than the one from Eq.~(\ref{annihcore}). 

However, if SHDM particles are fermions like in the case of
neutralinos, there is quite different effect which stops the core
contraction at a much larger radius. This effect is the pressure of
a degenerate Fermi gas. The maximum density of the core and hence the 
radius of the  core can be derived from the equality of the Fermi momentum of a 
degenerate gas and the virial momentum of the constituent particles at
the radius of the core $r=r_c$. 
\begin{equation}
p_F=(3\pi^2)^{1/3}(\rho_c/m_\chi)^{1/3} =m_\chi V_c \,,
\end{equation}
where  $V_c= \sqrt{GM_c/r_c}$ is the virial velocity at the boundary 
of the core, and $M_c=(4\pi/3)\rho_c r_c^3$ is the mass of the core. 
For the density profile $\rho(r) \propto r^{-2}$, the virial velocity is
the same at all $r$ and we can take it for the whole clump with mass
$M$ and radius $R$. Using core radius from $x_c=(\bar\rho/3\rho_c)^{1/2}$
we obtain 
\be
x_c^2=\pi^2 \frac{\bar\rho}{m_\chi^4}\left (\frac{GM}{R}\right )^{-3/2}.
\label{x_c}
\ee
For a clump with mass $M \sim 1\times 10^5$~g, 
mean density $\bar\rho \sim 3\times 10^3$~g/cm$^3$ and $R \sim 3$~cm, 
the radius of the core is $x_c \sim 1\times 10^{-11}$. 

In our calculations below  for superheavy neutralinos, we will use 
for radius of the core Eq.~(\ref{x_c}). 

\subsection{Properties of superdense clumps from SHDM particles and
numerical examples}
\label{properties}

The detectability of the annihilation signal from DM clumps
is determined by the following parameters: The mass $m_\chi$ of the
superheavy neutralino and its mixing parameters, the 
density profile, e.g.\ an isothermal profile  
$\rho (r)=\rho_c (r/R_c)^{-2}$ in case of a 'gravithermal catastrophe', 
the dimensionless radius of the core 
$x_c=R_c/R=(\bar\rho/3\rho_c)^{1/2}$ and the maximal density $\rho_c$. 
The important parameter, the mean density of a clump $\bar\rho$, is
found from the evolution of a primordial perturbation with initial
amplitude $\delta_H$ for a clump of given mass $M$. The mean densities 
are shown in Fig.~\ref{figmulti} for different $M$ and $\delta_H$. 
The maximum density, and respectively the radius of the core, are found as
described above. 

As particular examples we consider three sets of clump parameters
which we shall use in the next section for the calculation of the 
annihilation signal from superdense clumps.
We discuss first an optimistic example for a bino with mass 
$m_\chi=10^{11}$\,GeV. We consider clumps with $M\simeq10^5$~g formed
from fluctuations with $\delta_H\simeq0.09$, marked by a star in 
Fig.~\ref{figmulti}.  The density profile of such clumps is
$\rho(r)\propto r^{-2}$ (we choose the isothermal value, which is close to the 
analytical and numerical results $\beta\simeq1.7-2$) and the initial core 
radius $x_{c,1}\sim0.01$. Such clumps have mean density 
$\bar\rho\simeq1.3\times10^{3}$~g~cm$^{-3}$, radius
$R\simeq2.6$~cm, and virial velocity $v\simeq0.05$~cm/s. For these
parameters, the evolution time of the initial core is $t_e\simeq 0.4t_0$. 
Fermi degeneracy mainly restricts the central density in this case, and
the new core radius is $R_{c,2}\simeq2.9\times10^{-11}$~cm ($x_{c,2}\simeq1.1\times10^{-11}$).

The parameters of a more typical example (marked by a cross in 
Fig.~\ref{figmulti}) are $M\simeq10^{15}$~g  and $\delta_H\simeq0.07$. 
These clumps have the same density profile and initial core radius,
$\rho(r)\propto r^{-2}$ and $x_{c,1}\sim0.01$. But now  $\bar\rho\simeq6.3\times10^{-11}$~g~cm$^{-3}$, $R\simeq1.6\times10^8$~cm, 
$v\simeq0.65$~cm/s, and no singularity forms in the clumps.

In the case of a higgsino with the same mass, $m_\chi=10^{11}$\,GeV, an 
optimistic choice of parameters corresponds to clumps with 
$M=M_d\simeq6\times10^{21}$~g.  The density profile of such clumps is
$\rho(r)\propto r^{-2}$ and the core radius $x_{c,1}\sim0.01$. Also in this case 
no gravithermal catastrophe develops.
These clumps have $\bar\rho\simeq10^{-6}$~g~cm$^{-3}$, 
$R\simeq10^9$~cm, and $v\simeq600$~cm/s. 

\begin{figure}[t]
\begin{center}
\includegraphics[angle=0,width=0.45\textwidth]{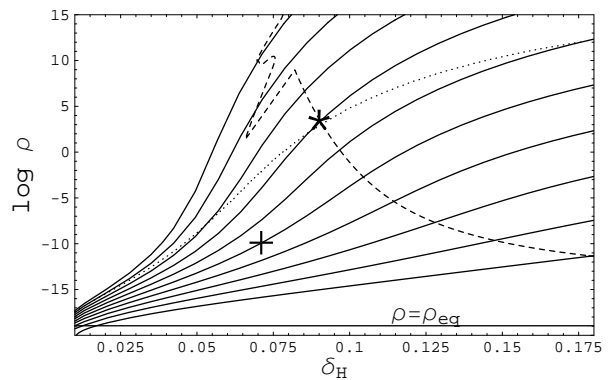}
\end{center}
\caption{The  mean density $\rho$ (in g~cm$^{-3}$) of DM clumps as function 
of the perturbation $\delta_{\rm H}$ in the radiation density on the horizon 
scale; solid lines from top to bottom are for for clump masses $M=10^{-10}$, 
$10^{-5}$, $1$, $10^{5}$, $10^{10}$, $10^{15}$, $10^{20}$, $10^{25}$, $10^{30}$,
$10^{35}$~g. The dashed line is the bound on the clump density from 
primordial black holes overproduction with threshold $\delta_c=0.7$.
The time of two-body gravitational relaxation
inside the clump cores is less then the age of the Universe for
clumps above the dotted lines for $m_\chi=10^{11}$~GeV. The star
marks favourable parameters for annihilations, and the
cross marks a typical example considered for comparison.
\label{figmulti}}
\end{figure}
 
\section{Annihilation}
\label{sec:detect}

\subsection{Annihilation rate}
\label{annihs}

The annihilation rate $\dot N_{\rm ann}$ of neutralinos in a single clump
is
\begin{equation}
 \dot N_{\rm ann} = \frac{1}{2} \int_{0}^{R} \!\!\! 4\pi r^2dr \: n^2(r)
 \langle\sigma_{\rm ann} v\rangle  =
\frac{3}{8\pi} \frac{\langle\sigma_{\rm ann} v\rangle}{m_\chi^2}
\frac{M^2}{R^3}S \,,
\end{equation}
where $n(r)=\rho(r)/m_\chi$ is the number density 
of neutralinos as function of the distance to the core of the cloud. 
The function $S$ was determined in
Ref.~\cite{bde03} and depends on the distribution of DM in the clump. 
In particular, the function is $S=1$ for the simplest case of an uniform 
clump and $S\simeq 4/(9x_c)$ for an isothermal profile $\rho\propto r^{-2}$ 
with a small core size, $x_c\ll 1$. 

The resulting flux $I_i$ of particles of type $i=N,\gamma,\nu$ 
from DM annihilations summed over all DM clumps in the Galactic halo 
is given by
\be
 I_i(E) = 
\frac{1}{2}\,\dot N_{\rm ann}  {\cal H} \:\frac{1}{m_\chi}\,\frac{dN_i}{dx} \,,
\ee
where $dN_i/dx$ is the differential number of particles of type $i$ produced
per annihilation with energy $E=xm_\chi$. We calculate these spectra  as 
described in Ref.~\cite{BKmc,FF} for the case of a non-supersymmetric evolution 
of the fragmentation functions $dN_i/dE$. 

The function $ {\cal H}$ contains the 
information about the smooth DM distribution in the halo, 
\be
 {\cal H} = \int_0^{\pi} d\zeta\sin\zeta
 \int_{0}^{r_{\rm max}(\zeta)} ds \:\frac{\xi\rho_h[r(s,\zeta)]}{M} \,,
\ee
where $\xi$ is the fraction of DM in form of neutralino clumps,
$n_{\rm cl}(r)=\xi\rho_{\rm h}/M$ is the number density of clumps
at distance $s$ from the Sun along the line-of-sight (l.o.s.),
and $\zeta$ is the angle between the direction in the sky and the
galactic center (GC). Finally,
$r_{\max}= (R_{\rm H}^2-r_{\odot}^2\sin^2\zeta)^{1/2} + r_{\odot}\cos\zeta$ is
the distance to the border of the DM halo of  radius $r_h$ and 
$r_{\odot}=8.5$~kpc is the distance of the Sun to the Galactic center.

As distributions of the DM in the galactic halo we use the 
Navarro-Frenk-White profile~\cite{NFW}, 
\be
 \rho_{\rm h}(R) = \frac{\rho_0}{(R/R_s)^\alpha(1+R/R_s)^{2}} \,,
\label{n_X}
\ee
with $\alpha=1$, scale radius $R_s=20$\,kpc, $R_h=200$\,kpc as the size of 
the DM halo and $\rho_{\rm h}(r_\odot)=0.3$\,GeV/cm$^3$ as the DM density at the 
position of the Sun.

Annihilations may proceed in the clumps with different parameters because
the a priori unknown position and height of the putative spike in the spectrum 
of perturbation. The formation and the properties 
of superdense clumps were considered in Paper~I~\cite{pI}. Following
the formalism of Ref.~\cite{pI}, the clumps density follows as shown in 
Fig.~\ref{figmulti}. For illustration we use in our calculations the three 
sets of clumps parameters presented in the previous section.

\subsection{Cross section}

Analytical approximations for the annihilation cross section of a neutralino
valid in the limit $M_{\rm SUSY}\gg m_Z$  were presented in Ref.~\cite{BKS}
using lowest order perturbation theory. We use as annihilation cross section
for both binos and higgsino the subprocesses $\chi\chi\to ZZ$
and $\chi\chi\to W^+W^-$ in the case of a higgsino,
\be \langle\sigma_{\rm ann} v\rangle \simeq 2\times 10^{-42}
m_{11}^{-2} \mbox{~cm}^3~s^{-1} \,.
\ee
Annihilations into $Zh^0$ and $ZA$ can increase the cross-section,
depending on the values of the SUSY breaking parameters.

Since the relative velocities of 
SHLSPs in DM clumps are small, $\beta=v\ll 1$,  factors $g^2/\beta$
or $\ln(g^2/\beta)$ can lead to a break-down of perturbation theory.
This effect, first studied by Sommerfeld for Coulomb interactions, 
was generalized in  Ref.~\cite{Hisano} to the exchange of massive 
non-abelian gauge bosons relevant for neutralino annihilations. In the
case of a wino or higgsino, the small mass splitting 
$\delta m=m_{\chi}-m_{\chi^\pm_1}$ between the lightest neutralino and the
lightest chargino means that the charginos produced in 
$\chi\chi\to\chi_1^+\chi_1^-$ have the same small velocity as the 
neutralinos. Therefore multiple photon, $Z$ and $W^\pm$ exchange between the 
charginos becomes important. 

The resulting enhancement of the annihilation cross
section can be calculated non-relativistically and is, neglecting bound-state
effects, characterized by two parameters~\cite{CST}: The ratio 
$\eps=m_W/m_\chi$ determines, if the annihilation proceeds in the Coulomb 
($\eps\ll 1$) or in the Yukawa ($\eps\gg 1$) regime, while the ratio 
$x=g_{\rm eff}^2/\beta$ of the squared effective
coupling constant and the velocity determines, if factors $g_{\rm eff}^2/\beta$
or $\ln(g_{\rm eff}^2/\beta)$ lead to a break-down of perturbation theory.
Here, the effective coupling constant $g_{\rm eff}$ includes
all pre-factors in front of the Yukawa potential, as e.g.\
mixing factors.

The Sommerfeld factor ${\cal R} $ as ratio of the perturbative and 
non-perturbative annihilation cross section is given in the Coulomb case by
\be \label{SF}
   {\cal R} = \frac{\sigma_{\rm np}}{\sigma_{\rm pert}}\sim 
  \frac{\eta}{1-\exp(-\eta)}
\ee
with $\eta = \pm g_{\rm eff}^2/(2\beta)$.
Using in the case of a higgsino the parameters for the optimistic example 
given in Sec.~\ref{annihs}, we find a strong enhancement of the
annihilation rate, 
${\cal R}\sim 10^{10}$.
On the other hand, higgsino are relatively tightly coupled to the thermal
plasma, leading to a relatively large value of $M_d$.   Thus there is no 
gravithermal catastrophe for higgsino clump and we use $S=1$.

For a bino, the Sommerfeld effect is not effective, ${\cal R}=1$. Using
the above parameters for the optimistic example, we find 
$S\simeq 4/(9x_{c,2})\sim 4\times 10^{10}$.  Hence, annihilations of binos in 
clumps formed during the RD epoch can be enhanced by the factor
$S\sim 4\times 10^{10}$  compared to the annihilation signal computed for a 
smooth DM distribution inside clumps.

\begin{figure}[t]
\begin{center}
\includegraphics[angle=0,width=0.45\textwidth]{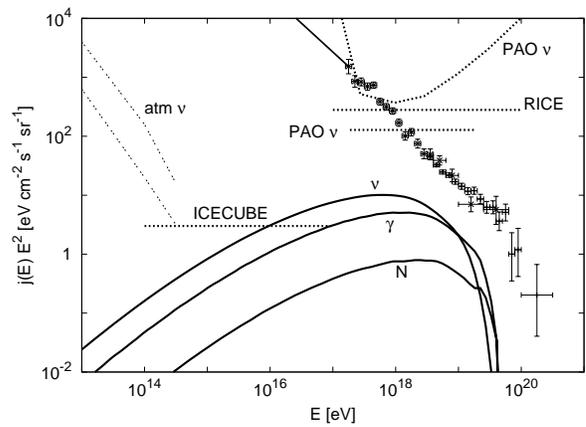}
\end{center}
\caption{
The maximal fluxes $I_i(E)$ of photons, nucleons and neutrinos from 
neutralino annihilations in Galactic halo together with experimental data
for a neutralino with 
$10^{11}$\,GeV.
\label{flux}}
\end{figure}
The maximal fluxes of photons, nucleons and neutrinos from a SHLSP
with $m_\chi=10^{11}$\,GeV allowed by cosmic ray 
data are shown in Fig.~\ref{flux} together with
upper limits for the neutrino fluxes from the Pierre Auger
experiment~\cite{PAO_nu} and RICE as well as the expected 90\% 
sensitivity after five years data-taking of ICECUBE~\cite{ice}.
For the values given in the optimistic example for a bino,
the flux shown in Fig.~\ref{flux} has been rescaled by the factor $10^{-14}$, 
while for a higgsino the flux was rescaled by $10^{-5}$. 
Thus the fluxes of secondaries overshoot the experimental data by 
many orders of magnitude for the most optimistic scenarios. This offers 
the potential to test different DM masses as well as different scenarios 
for the formation of superdense clumps. At present,
annihilations of SHLSPS are mainly restricted by experimental limits 
on the photon fraction~\cite{photon} and a galactic anisotropy 
of the  UHECR flux~\cite{search}. However, in the future neutrino searches 
at lower energies by a km$^3$ neutrino telescope as ICECUBE may become 
competitive.

\section{Conclusions}
\label{sec:conc}

We have studied the properties of SHDM clumps in two different cosmological 
scenarios, one with power-law and one with spiky density perturbations.   
As superheavy DM particles we have considered the superheavy neutralino, which 
properties were studied within superheavy supersymmetry in Ref.~\cite{BKS}.

For standard power-law fluctuations, SHDM clumps are formed in the DM
dominated epoch in hierarchical structures, when a small clump is
hosted by a bigger clump, which in turn is submerged into an even
bigger one, etc. Small clumps are tidally disrupted in such structures
and only a small fraction of the clumps survives. The surviving  clumps 
can be further destroyed by tidal interaction in the Galaxy. 

In contrast, clumps in  
spiky density perturbations are born in the RD dominated epoch, when 
hierarchical structures are not yet formed. They evolve as the single 
isolated objects without tidal destruction and merging. 

For the standard cosmological scenario the density of SHDM particles 
in a clump is low and the annihilation signal is weak, caused by the too small 
annihilation cross section. These clumps can be detected only when 
clumps are passing by gravitational wave detectors.  

In superdense clumps, an isothermal profile may continue up to very small 
radius of the new core, if the clump went through the ``gravithermal 
catastrophe''. The density of particles in the core and nearby is very 
large and this enhances the annihilation signal. Another reason for the
increase of the annihilation signal in comparison with clumps of 
EW mass particles is the small $M_{\rm min}$ in the mass distribution of the 
clumps. The cutoff $M_{\rm min}$ is smaller for a bino, but a higgsino gains 
from the Sommerfeld factor, which increases the annihilation cross section.    

As a result, we found that the annihilation
rate of stable superheavy neutralinos may be large enough to be detectable,
if primordial density perturbations are spiky. Hence the search for
photons or a galactic anisotropy in UHECRs~\cite{search} as well as
the search for UHE neutrinos offers not only 
the potential to identify the DM candidate but also to learn about the 
inflationary potential. 

\vskip0.4cm
\begin{acknowledgments}
VB is grateful to the A.~Salam International Centre for Theoretical Physics
for hospitality during the work on this paper and to A.~Smirnov for valuable
discussions. VD and YuE were supported in part by the Russian Federal Agency 
for Science and Innovation under state contract 02.740.11.5092 and by the 
grants of the Leading scientific schools 959.2008.2 and 438.2008.2.
\end{acknowledgments}



\begin{thebibliography}{99}



\bibitem{DMreviews}
G.~Bertone, D.~Hooper and J.~Silk,
Phys.\ Rept.\  {\bf 405}, 279 (2005)
[hep-ph/0404175].
L.~Bergstrom,
to appear in New J. of Phys,.
  arXiv:0903.4849 [hep-ph].


\bibitem{WMAP5}
J.~Dunkley {\it et al.}  [WMAP Collaboration],
Astrophys.\ J.\ Suppl.\  {\bf 180}, 306 (2009)
[arXiv:0803.0586 [astro-ph]].


\bibitem{GKH}
K.~Griest and M.~Kamionkowski,
Phys.\ Rev.\ Lett.\  {\bf 64}, 615 (1990);
L.~Hui,
Phys.\ Rev.\ Lett.\  {\bf 86}, 3467 (2001)
[astro-ph/0102349].

\bibitem{BKV}
V.~Berezinsky, M.~Kachelrie{\ss} and A.~Vilenkin,
Phys.\ Rev.\ Lett.\  {\bf 79}, 4302 (1997)
[astro-ph/9708217].

\bibitem{KR}
V.~A.~Kuzmin and V.~A.~Rubakov,
Phys.\ Atom.\ Nucl.\  {\bf 61}, 1028 (1998)
[Yad.\ Fiz.\  {\bf 61}, 1122 (1998)]
[astro-ph/9709187].


\bibitem{grav}
D.~J.~H.~Chung, E.~W.~Kolb and A.~Riotto,
Phys.\ Rev.\ D {\bf 59}, 023501 (1999)
[hep-ph/9802238];
V.~Kuzmin and I.~Tkachev,
JETP Lett.\  {\bf 68}, 271 (1998)
[hep-ph/9802304];
see also
D.~H.~Lyth and D.~Roberts,
Phys.\ Rev.\ D {\bf 57}, 7120 (1998)
[hep-ph/9609441].


\bibitem{BKmc}
V.~ Berezinsky and M.~ Kachelrie{\ss},
Phys.Rev. {\bf D63}, 034007 (2001) 
[arXiv:hep-ph/0009053]

\bibitem{BlasiDickKolb02} 
P.~Blasi, R.~Dick, E.~W.~Kolb,
Astropart. \ Phys.\  {\bf 18}, 57 (2002)
[arXiv:astro-ph/0105232v3].

\bibitem{kt} 
E.~W.~Kolb and I.~I.~Tkachev,
Phys.\ Rev.\  D {\bf 50}, 769 (1994)
[arXiv:astro-ph/9403011].

\bibitem{SS}
  P.~Scott and S.~Sivertsson,
Phys.\ Rev.\ Lett.\  {\bf 103}, 211301 (2009)
[arXiv:0908.4082 [astro-ph.CO]].


\bibitem{BKS}
V.~Berezinsky, M.~Kachelrie{\ss} and M.~Aa.~Solberg,
Phys.\ Rev.\  D {\bf 78}, 123535 (2008)
[arXiv:0810.3012 [hep-ph]].

\bibitem{bde06} 
V. Berezinsky, V. Dokuchaev and Yu.~Eroshenko,
Phys.\ Rev.\  D {\bf 73}, 063504 (2006)
[arXiv:astro-ph/0511494].

\bibitem{cutweak}
C.~Schmid, D.~J.~Schwarz and P.~Widerin,
  Phys.\ Rev.\  D {\bf 59}, 043517 (1999)
  [arXiv:astro-ph/9807257];
A.~M.~Green, S.~Hofmann and D.~J.~Schwarz,
  JCAP {\bf 0508}, 003 (2005)
  [arXiv:astro-ph/0503387];
  S.~Profumo, K.~Sigurdson and M.~Kamionkowski, 
Phys. \ Rev. \ Lett. {\bf 97}, 031301 (2006)
[arXiv:astro-ph/0603373v1];
  E.~Bertschinger,
  Phys.\ Rev.\  D {\bf 74}, 063509 (2006)
  [arXiv:astro-ph/0607319].


\bibitem{GelGon08}
G.~B.~Gelmini and P.~Gondolo,
JCAP {\bf 0810}, 002 (2008)
[arXiv:0803.2349 [astro-ph]].



\bibitem{pI}
V.~Berezinsky, V.~Dokuchaev, Yu.~Eroshenko, M.~Kachelrie{\ss} 
and M.~A.~Solberg,
[arXiv:1002.3444 [astro-ph.CO]]
(Paper I).

\bibitem{LoeZal05}
A.~Loeb and M.~Zaldarriaga,
Phys.\ Rev.\  D {\bf 71}, 103520 (2005)
[arXiv:astro-ph/0504112].

\bibitem{ind} T. Padmanabhan and K. Subramanian, Astrophys.\ J.\ {\bf
417}, 3 (1993).


\bibitem{ps74} W. H. Press and P. Schechter, Astrophys.\ J.\ {\bf 187},
425 (1974).

\bibitem{cole} C. Lacey, S. Cole, Mon.\ Not.\ R.\ Astron.\ Soc.\ {\bf 262},
627 (1993).

\bibitem{bde08} 
V.~Berezinsky, V.~Dokuchaev and Y.~Eroshenko,
Phys.\ Rev.\  D {\bf 77}, 083519 (2008)
[arXiv:0712.3499 [astro-ph]].

\bibitem{bde03} 
V.~Berezinsky, V.~Dokuchaev and Y.~Eroshenko,
Phys.\ Rev.\  D {\bf 68}, 103003 (2003)
[arXiv:astro-ph/0301551].

\bibitem{DieMooSta05} 
J.~Diemand, B.~Moore and J.~Stadel,
Nature {\bf 433}, 389 (2005)
[arXiv:astro-ph/0501589].

\bibitem{ufn1}
 A. V. Gurevich and K. P. Zybin,
Sov.\ Phys.\ JETP {\bf 67}, 1 (1988); Sov.\ Phys.\ JETP {\bf
67}, 1957 (1988); Sov.\ Phys.\ Usp.\ {\bf 165}, 723 (1995).

\bibitem{IshMakEbi}T. Ishiyama, J. Makino and T. Ebisuzaki (to be published).

\bibitem{Spit}
L.~Spitzer and W.C.~Saslaw, Astrophys.\ J.\ {\bf 143}, 400
(1966).

\bibitem{koch00}
C.~S.~Kochanek and M.~J.~White,
Astrophys.\ J.\  {\bf 543}, 514 (2000)
[arXiv:astro-ph/0003483].


\bibitem{bergur} 
V.~S.~Berezinsky, A.~V.~Gurevich and K.~P.~Zybin,
  Phys.\ Lett.\  B {\bf 294}, 221 (1992).


\bibitem{berez96} 
V.~Berezinsky, A.~Bottino and G.~Mignola,
  Phys.\ Lett.\  B {\bf 391}, 355 (1997)
  [arXiv:astro-ph/9610060].


\bibitem{FF}
R.~Aloisio, V.~Berezinsky and M.~Kachelrie\ss,
  Phys.\ Rev.\  D {\bf 69}, 094023 (2004)
  [arXiv:hep-ph/0307279].

\bibitem{Hisano}
J.~Hisano, S.~Matsumoto and M.~M.~Nojiri,
  Phys.\ Rev.\  D {\bf 67}, 075014 (2003)
  [arXiv:hep-ph/0212022];
J.~Hisano, S.~Matsumoto, M.~M.~Nojiri and O.~Saito,
  Phys.\ Rev.\  D {\bf 71}, 015007 (2005)
  [arXiv:hep-ph/0407168];
J.~Hisano, S.~Matsumoto, M.~M.~Nojiri and O.~Saito,
  Phys.\ Rev.\  D {\bf 71}, 063528 (2005)
  [arXiv:hep-ph/0412403].


\bibitem{CST}
M.~Cirelli, A.~Strumia and M.~Tamburini,
Nucl.\ Phys.\  B {\bf 787}, 152 (2007)
[arXiv:0706.4071 [hep-ph]].

\bibitem{NFW} 
J.~F.~Navarro, C.~S.~Frenk and S.~D.~M.~White,
  Astrophys.\ J.\  {\bf 462}, 563 (1996)
  [arXiv:astro-ph/9508025].

\bibitem{PAO_nu}
  J.~Abraham {\it et al.}  [The Pierre Auger Collaboration],
  Phys.\ Rev.\ Lett.\  {\bf 100}, 211101 (2008)
  [arXiv:0712.1909 [astro-ph]].


\bibitem{ice}
  J.~Ahrens {\it et al.}  [IceCube Collaboration],
  Astropart.\ Phys.\  {\bf 20}, 507 (2004)
  [arXiv:astro-ph/0305196].

\bibitem{photon}
G.~I.~Rubtsov {\it et al.},
  Phys.\ Rev.\  D {\bf 73}, 063009 (2006)
  [arXiv:astro-ph/0601449];
J.~Abraham {\it et al.}  [Pierre Auger Collaboration],
  Astropart.\ Phys.\  {\bf 27}, 155 (2007)
  [arXiv:astro-ph/0606619],
J.~Abraham {\it et al.}  [Pierre Auger Collaboration],
  ibid.\ {\bf 29}, 243 (2008)
  [arXiv:0712.1147 [astro-ph]].


\bibitem{search}
R.~Aloisio, V.~Berezinsky and M.~Kachelrie\ss,
  Phys.\ Rev.\  D {\bf 74}, 023516 (2006)
  [arXiv:astro-ph/0604311];
R.~Aloisio and F.~Tortorici,
  Astropart.\ Phys.\  {\bf 29}, 307 (2008)
  [arXiv:0706.3196 [astro-ph]].
For a recent review see
M.~Kachelrie\ss,
  arXiv:0810.3017 [astro-ph].
  
  
  
 
\end{thebibliography}
\end{document}